\documentclass[showkeys,showpacs,prd,aps,floats,eqsecnum,preprint]{revtex4}

\begin{document}

\title{Conserved currents in gravitational models with 
quasi-invariant Lagrangians: Application to teleparallel gravity}

\author{Yuri N.~Obukhov}
\email{yo@ift.unesp.br}
\affiliation{Instituto de F\'{\i}sica Te\'orica,
Universidade Estadual Paulista \\
Rua Pamplona 145,  01405-900 S\~ao Paulo, Brazil}
\affiliation{Department of Theoretical Physics,
Moscow State University, 117234
Moscow, Russia}

\author{Guillermo F.~Rubilar}
\email{grubilar@udec.cl}
\affiliation{Instituto de F\'{\i}sica Te\'orica,
Universidade Estadual Paulista \\
Rua Pamplona 145,  01405-900 S\~ao Paulo, Brazil}
\affiliation{Departamento de F{\'{\i}}sica,
Universidad de Concepci\'on, Casilla 160-C, Concepci\'on, Chile}

\author{J.~G.~Pereira}
\email{jpereira@ift.unesp.br}
\affiliation{Instituto de F\'{\i}sica Te\'orica,
Universidade Estadual Paulista \\
Rua Pamplona 145,  01405-900 S\~ao Paulo, Brazil}

\begin{abstract}
Conservation laws in gravitational theories with diffeomorphism
and local Lorentz symmetry are studied. Main attention is paid to
the construction of conserved currents and charges associated 
with an arbitrary vector field that generates a diffeomorphism on
the spacetime. We further generalize previous results for the case
of gravitational models described by quasi-invariant Lagrangians, 
that is, Lagrangians that change by a total derivative under the 
action of the local Lorentz group. The general formalism is then 
applied to the teleparallel models, for which the energy and the 
angular momentum of a Kerr black hole are calculated. The subsequent
analysis of the results obtained demonstrates the importance of 
the choice of the frame.
\end{abstract}


\maketitle

\section{Introduction}

In this paper, we continue the study of conservation laws in
gravitational theories with diffeomorphism and local
Lorentz symmetries. The previous results, as well as an overview
of the earlier literature can be found in \cite{conserved,invar},
see also the extensive reference list in \cite{Szabados}. More
specifically, we will be interested in the conserved currents and
charges associated with vector fields that generate arbitrary
diffeomorphisms on the spacetime manifold.

Most gravitational models are normally invariant
under both spacetime diffeomorphisms and local Lorentz transformations
of the frames. However, the Lagrangian may
be shifted by a total derivative term (equivalently, by a boundary
term). This actually happens in many gravitational field models: 
(i) when one adds a noninvariant boundary term to the original 
Lagrangian, (ii) when the gravitational dynamics is described in the 
purely tetrad framework \cite{conserved}, (iii) when the Lagrangian 
includes topological terms (e.g., of the Chern-Simons type 
\cite{nieh1,nieh2,CS}), typically in 3 and 5 dimensions, for example 
\cite{miebaek,kawai,cs,mil,3d,Bak94}.

For models with Lorentz-invariant Lagrangians, we have shown in 
\cite{invar} that it is possible to define \textit{invariant} conserved 
currents for every vector field $\xi$. These conserved currents do not 
depend on the coordinate system or the tetrad frame used to compute 
them. They depend only on the field configuration and on the choice of the 
vector field $\xi$. The interpretation of $\xi$ is an important geometrical
and physical issue. For example, when $\xi$ is timelike, the corresponding
charge has the meaning of the energy of the gravitating system with respect 
to an observer moving along the integral lines of $\xi$, with 4-velocity
$u=\xi/|\xi|$, cf. \cite{KLB06}. In this way, the dependence of the conserved
charges on $\xi$ describes the usual dependence of the energy of a system on
the choice and on the dynamics of a physical observer. Furthermore, let us
recall that under certain conditions discussed in \cite{invar}, the 
\textit{matter} current  $J^{\rm mat}[\xi]= \xi^iT_i{}^j\partial_j\rfloor
\eta$ is conserved, with $T_i{}^j$ being the energy-momentum tensor of matter,
and $\eta$ the volume 4-form. The corresponding invariant charge ${\cal 
Q}^{\rm mat}[\xi]=\int_S J^{\rm mat}[\xi]$ is, therefore, the integral of 
the {\it projection} of the energy-momentum along the vector $\xi$. The
charge ${\cal Q}^{\rm mat}[\xi]$ reduces to the usual expression $\int_S
T_0{}^0\sqrt{-g}\, dx^1\wedge dx^2\wedge dx^3$ in coordinates adapted to $\xi$
such that $\xi=\partial_0$, and the hypersurface $S$ is defined by $x^0=$
constant. 
This general idea of defining the energy of a system as a scalar (i.e., 
invariant) depending on some vector field is a generalization of the 
well-known construction for point particles, see, for instance, Sect. 2.8 
(and in particular Eq. (2.29)) of \cite{MTW73}. 

In the first part of this paper we generalize the formalism 
developed in \cite{invar}, and define conserved currents and charges 
for models with ``quasi-invariant" Lagrangians (that is, Lagrangians 
that change by a total derivative under local Lorentz transformations). 
In the second part we apply this formalism to teleparallel gravity.
As is well known, in addition to the geometric framework of
general relativity, gravitation can also be described in terms
of a gauge theory. In fact, teleparallel gravity corresponds to
a gauge theory for the translation group \cite{Gronwald97,HSh79,Itin}. 
In this theory, instead of curvature, torsion represents the gravitational
field. In spite of this fundamental difference, teleparallel gravity 
is found to be equivalent to general relativity. This means essentially that 
the gravitational interaction can be described {\em alternatively} 
in terms of curvature, as is usually done in general relativity, or 
in terms of torsion, as in teleparallel gravity.

One may wonder why gravitation has two different descriptions.
Such a dualism is related to a peculiar property of gravitation,
called universality. As is well known, at least at the classical
level, particles with different masses and different
compositions feel it in such a way that all of them acquire the
same acceleration and, given the same initial conditions, follow
the same path. Such universality of response is one of the most
fundamental characteristic of the gravitational interaction. It
is unique, peculiar to gravitation: no other basic interaction
of Nature has it. And it is exactly this property that makes a 
geometrized formulation of gravitation possible in
addition to the gauge description of teleparallel gravity. As
the sole universal interaction, it is the only one to allow a
geometrical interpretation, and two alternative descriptions.
One may also wonder why a gauge theory for the translation
group, and not for other spacetime group. The reason for this is
related to the source of gravitation, that is, energy and
momentum. As is well known from Noether's theorem,
these quantities are conserved provided the physical system is
invariant under spacetime translations. It is then natural to
expect that the gravitational field be associated to the
translation group. This is quite similar to the electromagnetic
field, whose source --- the electric four-current --- is
conserved due to invariance of the theory under transformations
of the unitary group $U(1)$, the gauge group of Maxwell's theory.

Our general notations are the same as in \cite{HMMN95}. In
particular, we use the Latin indices $i,j,\dots$ for local
holonomic spacetime coordinates and the Greek indices $\alpha,\beta,
\dots$ label (co)frame components. Particular frame components are
denoted by hats, $\hat 0, \hat 1$, etc. As usual, the exterior
product is denoted by $\wedge$, while the interior product of a
vector $\xi$ and a $p$-form $\Psi$ is denoted by $\xi\rfloor\Psi$.
The vector basis dual to the frame 1-forms $\vartheta^\alpha$ is
denoted by $e_\alpha$ and they satisfy 
$e_\alpha\rfloor\vartheta^\beta
=\delta^\beta_\alpha$. Using local coordinates $x^i$, we have
$\vartheta^\alpha=h^\alpha_idx^i$ and 
$e_\alpha=h^i_\alpha\partial_i$.
We define the volume $n$-form by $\eta:=\vartheta^{\hat{0}}\wedge
\cdots\wedge\vartheta^{\hat{n}}$. Furthermore, with the help of
the interior product we define $\eta_{\alpha}:=e_\alpha
\rfloor\eta$, $\eta_{\alpha\beta}:=e_\beta\rfloor\eta_\alpha$,
$\eta_{\alpha\beta\gamma}:= e_\gamma\rfloor\eta_{\alpha\beta}$, 
etc., which are bases for $(n-1)$-, $(n-2)$- and $(n-3)$-forms, 
etc., respectively. Finally, $\eta_{\alpha_1\cdots \alpha_n} =
e_{\alpha_n}\rfloor\eta_{\alpha_1\cdots\alpha_{n-1}}$ is the
Levi-Civita tensor density. The $\eta$-forms satisfy the identities:
\begin{eqnarray}
\vartheta^\beta\wedge\eta_\alpha &=& \delta_\alpha^\beta\eta ,\\
\vartheta^\beta\wedge\eta_{\mu\nu} &=& \delta^\beta_\nu\eta_{\mu} -
\delta^\beta_\mu\eta_{\nu},\label{veta1}\\ \label{veta}
\vartheta^\beta\wedge\eta_{\alpha\mu\nu}&=&\delta^\beta_\alpha
\eta_{\mu\nu} + \delta^\beta_\mu\eta_{\nu\alpha} +
\delta^\beta_\nu\eta_{\alpha\mu},\\
\vartheta^\beta\wedge\eta_{\alpha\gamma\mu\nu}&=&\delta^\beta_\nu
\eta_{\alpha\gamma\mu} - \delta^\beta_\mu\eta_{\alpha\gamma\nu}
+ \delta^\beta_\gamma\eta_{\alpha\mu\nu} -
\delta^\beta_\alpha\eta_{\gamma\mu\nu},\label{veta2}
\end{eqnarray}
etc. The line element $ds^2 = 
g_{\alpha\beta}\,\vartheta^\alpha\otimes
\vartheta^\beta$ is defined by the spacetime metric 
$g_{\alpha\beta}$
of signature $(+,-,\cdots,-)$.

\section{The Lagrange-Noether machinery}\label{LN}

The gravitational field is described by the coframe 
$\vartheta^\alpha$ and the Lorentz connection $\Gamma_\alpha
{}^\beta$ 1-forms. The matter fields $\Psi^A$ can be 
Lorentz-covariant forms of an arbitrary rank $p$.

Let $V^{\rm tot} = V^{\rm tot}(\vartheta, d\vartheta, \Gamma, 
d\Gamma, \Psi^A,d\Psi^A)$ be an arbitrary Lagrangian. The 
total variation then formally reads as (cf. \cite{invar}):
\begin{eqnarray}\label{varV}
\delta V^{\rm tot} &=&\delta\vartheta^{\alpha}\wedge {\cal 
F}_\alpha + \delta\Gamma_\alpha{}^{\beta}\wedge {\cal F}^\alpha
{}_\beta + \delta\Psi^A\wedge {\cal F}_A \nonumber\\
&&-\,d\left(\delta\vartheta^{\alpha}\wedge {\cal H}_\alpha +
\delta\Gamma_{\alpha}{}^{\beta}\wedge {\cal H}^\alpha{}_\beta 
+ \delta\Psi^A\wedge {\cal H}_A\right).\label{var01}
\end{eqnarray}
Here as usual we introduce the generalized (translational, 
rotational, and matter) field momenta by
\begin{equation}
{\cal H}_{\alpha} := -\,{\frac{\partial V^{\rm tot}}{\partial
d\vartheta^{\alpha}}}\,,\qquad
{\cal H}^{\alpha}{}_{\beta} := -\,{\frac{\partial V^{\rm 
tot}}{\partial d\Gamma_{\alpha}
{}^{\beta}}},\qquad {\cal H}_A := -\,\frac{\partial V^{\rm 
tot}}{\partial d\Psi^A},\label{HH}
\end{equation}
whereas the variational derivatives w.r.t. the fields are defined by
\begin{eqnarray}
{\cal F}_\alpha &:=& \frac{\delta V^{\rm 
tot}}{\delta\vartheta^{\alpha}}
= \frac{\partial V^{\rm tot}}{\partial\vartheta^{\alpha}} 
- d{\cal H}_\alpha,\label{Fa}\\
{\cal F}^\alpha{}_\beta &:=& \frac{\delta V^{\rm tot}}
{\delta\Gamma_\alpha{}^\beta}
= \frac{\partial V^{\rm tot}}{\partial\Gamma_\alpha{}^\beta} - 
d{\cal H}^\alpha{}_\beta,\label{Fab} \\ 
{\cal F}_A &:=& \frac{\delta L}{\delta\Psi^A} = \frac{\partial 
V^{\rm tot}}{\partial\Psi^A} + (-1)^p\,d{\cal H}_A\label{EPsi}.
\end{eqnarray}

\subsection{Lagrangians for gravity and matter: field equations}

The total Lagrangian is a sum $V^{\rm tot} = V + L$ of the 
gravitational Lagrangian $V = V(\vartheta, d\vartheta, \Gamma, 
d\Gamma)$ and the matter Lagrangian $L = L(\Psi^A,d\Psi^A,\vartheta, 
d\vartheta, \Gamma, d\Gamma)$. The dependence of the latter on the 
derivatives of the gravitational potentials may arise for models 
with nonminimal coupling. Then one usually defines for the 
gravitational Lagrangian the derivatives:
\begin{eqnarray}
{\cal E}_\alpha &:=& {\frac{\delta V}{\delta\vartheta^{\alpha}}}
= -\,dH_\alpha + {\frac{\partial 
V}{\partial\vartheta^{\alpha}}},\label{Ea}\\
{\cal C}^\alpha{}_\beta &:=& {\frac{\delta V}{\delta\Gamma_\alpha
{}^\beta}} = -\,dH^\alpha{}_\beta + {\frac{\partial 
V}{\partial\Gamma_\alpha
{}^\beta}}.\label{Cab}
\end{eqnarray}
Here $H_\alpha$ and $H^\alpha{}_\beta$ are defined analogously
to (\ref{HH}), but for the gravitational Lagrangian $V$. These 
quantities describe the left-hand (geometric) sides of the 
gravitational field equations. Analogous variational 
derivatives of the matter Lagrangian define the energy-momentum 
and spin of matter:
\begin{equation}
\Sigma_\alpha := {\frac {\delta 
L}{\delta\vartheta^\alpha}},\qquad \tau^\alpha
{}_\beta := {\frac {\delta 
L}{\delta\Gamma_\alpha{}^\beta}}.\label{ST}
\end{equation}

The total system of coupled field equations then reads
\begin{eqnarray}
{\cal F}_\alpha &=& {\cal E}_\alpha + \Sigma_\alpha =0, 
\label{cFa}\\
{\cal F}^\alpha{}_\beta &=& {\cal C}^\alpha{}_\beta + 
\tau^\alpha{}_\beta =0,
\label{cFab}\\
{\cal F}_A &=& 0. \label{cFAA}
\end{eqnarray}

\subsection{Noether identities for the Lorentz symmetry}

Let us assume that for an infinitesimal Lorentz transformation,
\begin{equation}\label{deltaVG}
\delta\vartheta^\alpha = 
\varsigma\varepsilon^\alpha{}_\beta\,\vartheta^\beta,
\qquad \delta\Gamma_\beta{}^\alpha = - \varsigma 
D\varepsilon^\alpha{}_\beta,
\qquad \delta\Psi^A = \varsigma \varepsilon^\alpha{}_\beta
(\rho^\beta{}_\alpha)^A{}_B\Psi^B,
\end{equation}
with $\Lambda^\alpha{}_\beta = \delta^\alpha_\beta +
\varsigma\varepsilon^\alpha{}_\beta$,  $\varepsilon_{\alpha\beta}
= -\varepsilon_{\beta\alpha}$, and $\varsigma$ an infinitesimal 
constant parameter ($\rho^\beta{}_\alpha$ are the Lorentz 
generators for the matter fields), the Lagrangian is changed by 
a total derivative: 
\begin{equation}
\delta_\varepsilon V^{\rm tot} =
-\,d\left(\varsigma\varepsilon^\alpha{}_\beta\,v^\beta{}_\alpha
\right).
\end{equation}  
Here $v^\beta{}_\alpha$ is some $(n-1)$-form. In view of the skew
symmetry of the Lorentz parameters, it is also antisymmetric,
$v_{\alpha\beta} = - v_{\beta\alpha}$.

Then (\ref{var01}) yields straightforwardly
\begin{eqnarray}
\varepsilon^\alpha{}_\beta\left(\vartheta^\beta\wedge {\cal 
F}_\alpha +
D{\cal F}^\beta{}_\alpha + (\rho^\beta{}_\alpha)^A{}_B\Psi^B\wedge
{\cal F}_A \right) \nonumber\\
+ d\left[\varepsilon^\alpha{}_\beta\left(v^\beta
{}_\alpha - {\cal C}^\beta{}_\alpha - \vartheta^\beta\wedge 
{\cal H}_\alpha -
D{\cal H}^\beta{}_\alpha - 
(\rho^\beta{}_\alpha)^A{}_B\Psi^B\wedge {\cal
H}_A\right)\right] =0.
\end{eqnarray}
Notice that the covariant exterior derivative 
$D:=d+\rho^\beta{}_\alpha$ is used only as an abbreviation when 
acting on ${\cal H}^\beta{}_\alpha$ and $\varepsilon^\alpha{}_\beta$ in
(\ref{deltaVG}), since these quantities are not, in general, Lorentz-covariant
fields.

As a result, we find the two Noether identities (since
the transformation parameters $\varepsilon$ and their derivatives
$d\varepsilon$ are pointwise arbitrary):
\begin{eqnarray}
D{\cal F}_{\alpha\beta} + \vartheta_{[\alpha}\wedge
{\cal F}_{\beta]} + (\rho^\beta{}_\alpha
)^A{}_B\Psi^B\wedge {\cal F}_A &\equiv& 0,\label{Noe1}\\
v_{\alpha\beta} - {\cal F}_{\alpha\beta} - 
\vartheta_{[\alpha}\wedge
{\cal H}_{\beta]} - D{\cal H}_{\alpha\beta} - 
(\rho_{\alpha\beta})^A{}_B\Psi^B
\wedge {\cal H}_A &\equiv& 0.\label{Noe2}
\end{eqnarray}
The second relation is trivial in models with invariant 
Lagrangians.

\subsection{Noether identities for the diffeomorphism symmetry}

Let us derive the consequences of the assumed diffeomorphism 
invariance of $V^{\rm tot}$. Let $f$ be an arbitrary local diffeomorphism 
on the spacetime manifold. It acts with the pull-back map $f^\ast$ on all
the geometrical quantities, and the invariance of the theory means
that $V^{\rm tot}(f^\ast\vartheta,f^\ast d\vartheta,f^\ast \Gamma,
f^\ast d\Gamma, f^\ast \Psi,f^\ast d\Psi) = f^\ast(V^{\rm 
tot}(\vartheta, d\vartheta,\Gamma,d\Gamma,\Psi,d\Psi))$.
Consider an arbitrary vector field $\xi$ and the corresponding
local 1-parameter group of diffeomorphisms $f_t$ generated along
this vector field. Then, using $f_t$ in the above formula and
differentiating w.r.t. the parameter $t$, we find the identity
\begin{eqnarray}
(\ell_\xi\vartheta^\alpha)\wedge {\frac {\partial V^{\rm 
tot}}{\partial
\vartheta^\alpha}} + (\ell_\xi\Gamma_\alpha{}^\beta)\wedge {\frac
{\partial V^{\rm tot}}{\partial\Gamma_\alpha{}^\beta}} +
(\ell_\xi\Psi^A)\wedge
{\frac {\partial V^{\rm tot}}{\partial\Psi^A}} \nonumber\\
- (\ell_\xi d\vartheta^\alpha)\wedge {\cal H}_\alpha - (\ell_\xi d
\Gamma_\alpha{}^\beta)\wedge {\cal H}^\alpha{}_\beta - (\ell_\xi 
d\Psi^A)
\wedge {\cal H}_A = \ell_\xi V^{\rm tot}. \label{lieV1}
\end{eqnarray}
Moving the last term to the l.h.s., and using the Lie derivative 
(that is given on exterior forms by $\ell_\xi = d\xi\rfloor + \xi
\rfloor d$), we then find the identity $A + dB = 0$ with
\begin{eqnarray}
A &=& \xi^\alpha\left(-\,d{\cal F}_\alpha - 
e_\alpha\rfloor\Gamma_\gamma
{}^\beta\,d{\cal F}^\gamma{}_\beta + e_\alpha\rfloor 
d\vartheta^\beta
\wedge{\cal F}_\beta + e_\alpha\rfloor d\Gamma_\gamma{}^\beta\wedge
{\cal F}^\gamma{}_\beta\right. \nonumber\\
&&\left. +\,(-1)^pe_\alpha\rfloor\Psi^A\wedge d{\cal F}_A
+e_\alpha\rfloor d\Psi^A\wedge {\cal F}_A \right),\label{A1}\\
B &=& \xi^\alpha\left({\frac {\partial V^{\rm 
tot}}{\partial\vartheta^\alpha}}
+ e_\alpha\rfloor\Gamma_\beta{}^\gamma{\frac {\partial V^{\rm tot}}
{\partial\Gamma_\beta{}^\gamma}} + 
e_\alpha\rfloor\Psi^A\wedge{\frac
{\partial V^{\rm tot}}{\partial\Psi^A}} \right.\nonumber\\
&&\left. -\,e_\alpha\rfloor d\vartheta^\beta\wedge {\cal H}_\beta -
e_\alpha\rfloor
d\Gamma_\gamma{}^\beta\wedge {\cal H}^\gamma{}_\beta - 
e_\alpha\rfloor
d\Psi^A\wedge
{\cal H}_A - e_\alpha\rfloor V^{\rm tot}\right).\label{B}
\end{eqnarray}
Since the diffeomorphism invariance holds for {\it arbitrary}
vector fields $\xi$, then $A$ and $B$ must necessarily vanish, and 
we find the two Noether identities:
\begin{eqnarray}
d\left({\cal F}_\alpha + e_\alpha\rfloor\Gamma_\gamma{}^\beta\,{\cal
F}^\gamma{}_\beta + e_\alpha\rfloor\Psi^A\wedge{\cal F}_A\right)
&\equiv& \nonumber\\
(\ell_{e_\alpha}\vartheta^\beta)\wedge{\cal F}_\beta +
(\ell_{e_\alpha}\Gamma_\gamma{}^\beta)\wedge {\cal F}^\gamma{}_\beta
+ (\ell_{e_\alpha}\Psi^A)\wedge {\cal F}_A,\label{NoeD1}\\
{\frac {\partial V^{\rm tot}}{\partial\vartheta^\alpha}} +
e_\alpha\rfloor\Gamma_\beta{}^\gamma{\frac {\partial V^{\rm  tot}}
{\partial\Gamma_\beta{}^\gamma}} + e_\alpha\rfloor\Psi^A\wedge
{\frac {\partial V^{\rm tot}}{\partial\Psi^A}} &\equiv& \nonumber\\
e_\alpha\rfloor V^{\rm tot} + e_\alpha\rfloor d\vartheta^\beta\wedge
{\cal H}_\beta + e_\alpha\rfloor d\Gamma_\gamma{}^\beta\wedge {\cal
H}^\gamma{}_\beta + e_\alpha\rfloor d\Psi^A\wedge {\cal H}_A.\label{n02}
\end{eqnarray}

\section{Current associated with a vector field}

We define the current $(n-1)$-form by
\begin{equation}
J[\xi,\varepsilon] := \xi\rfloor V^{\rm tot} - 
\varepsilon^\alpha{}_\beta
\,v^\beta{}_\alpha + {\cal L}_{\{\xi,\varepsilon\}}\vartheta^\alpha
\wedge {\cal H}_\alpha +
{\cal L}_{\{\xi,\varepsilon\}}\Gamma_\alpha{}^\beta\wedge
{\cal H}^\alpha{}_\beta + {\cal L}_{\{\xi,\varepsilon\}}\Psi^A\wedge
{\cal H}_A.\label{Jdef}
\end{equation}
Here we introduced the ``generalized Lie derivatives" of the
gravitational fields by
\begin{eqnarray}
{\cal L}_{\{\xi,\varepsilon\}}\vartheta^\alpha &:=& 
\ell_\xi\vartheta^\alpha +
\varepsilon^\alpha{}_\beta\vartheta^\beta,\label{genLv}\\
{\cal L}_{\{\xi,\varepsilon\}}\Gamma_\alpha{}^\beta &:=&
\ell_\xi\Gamma_\alpha{}^\beta- 
D\varepsilon^\beta{}_\alpha,\label{genLg}\\
{\cal L}_{\{\xi,\varepsilon\}}\Psi^A &:=&  \ell_\xi\Psi^A
+\varepsilon^\alpha{}_\beta
(\rho^\beta{}_\alpha)^A{}_B\Psi^B,\label{genLPsi}
\end{eqnarray}
The current (\ref{Jdef}) satisfies the identity
\begin{equation}
dJ[\xi,\varepsilon] = {\cal 
L}_{\{\xi,\varepsilon\}}\vartheta^\alpha\wedge
{\cal F}_\alpha + {\cal 
L}_{\{\xi,\varepsilon\}}\Gamma_\alpha{}^\beta\wedge
{\cal F}^\alpha{}_\beta + {\cal L}_{\{\xi,\varepsilon\}}\Psi^A\wedge
{\cal F}_A.\label{dJ}
\end{equation}
This is just the total variation (\ref{var01}) in a different 
disguise.

Using Eqs.~(\ref{genLv})-(\ref{genLPsi}), the definitions 
(\ref{Ea}) and (\ref{Cab}), and the Noether identities (\ref{Noe2}) 
and (\ref{n02}), we can identically rewrite the current (\ref{Jdef}) 
in the form
\begin{equation}
J[\xi,\varepsilon] = d\left(\xi^\alpha {\cal H}_\alpha +
\Xi_\alpha{}^\beta[\xi,\varepsilon]\,{\cal H}^\alpha{}_\beta +
\xi\rfloor\Psi^A\wedge {\cal H}_A\right)
+ \xi^\alpha {\cal F}_\alpha + \Xi_\alpha{}^\beta[\xi,\varepsilon]
\,{\cal F}^\alpha{}_\beta + \xi\rfloor\Psi^A\wedge {\cal F}_A, 
\label{JdH}
\end{equation}
where we introduced the notation
\begin{equation}
\Xi_\alpha{}^\beta[\xi,\varepsilon] := 
\xi\rfloor\Gamma_\alpha{}^\beta -
\varepsilon^\beta{}_\alpha.
\end{equation}

\subsection{Gravitational current for Yano-Lie derivative}

There exist different choices of $\varepsilon^\alpha{}_\beta(\xi)$ 
for a given vector field $\xi$, that correspond to the use of 
different generalized Lie derivatives \cite{invar}. They give rise 
to different conserved currents and charges. The ``minimal''
choice \cite{invar}
\begin{equation}\label{Theta}
\varepsilon_{\alpha\beta} =  - \Theta_{\alpha\beta} = - 
e_{[\alpha}\rfloor\ell_\xi\vartheta_ {\beta]}
\end{equation}
defines the generalized Lie derivative (which we denote ${\cal 
L}_\xi$) in the sense of Yano \cite{Yano}. We will work with 
this choice for the rest of the paper.

Now, let us consider the case in which we take $V^{\rm tot}=V$, 
i.e. the Lagrangian of the gravitational field. In this case, 
the conserved current for an arbitrary quasi-invariant gravitational 
Lagrangian is found to be
\begin{equation}
{\cal J}^{\rm grav}[\xi] := \xi\rfloor V - 
\Theta^{\alpha\beta}v_{\alpha\beta}
+ {\cal L}_\xi\vartheta^\alpha\wedge H_\alpha + {\cal L}_\xi
\Gamma_\alpha{}^\beta\wedge H^\alpha{}_\beta.\label{JYdef}
\end{equation}
In accordance with the above derivations, it satisfies
\begin{equation}
d{\cal J}^{\rm grav}[\xi] = {\cal 
L}_\xi\vartheta^\alpha\wedge{\cal E}_\alpha
+ {\cal L}_\xi\Gamma_\alpha{}^\beta\wedge{\cal 
C}^\alpha{}_\beta,\label{dJY}
\end{equation}
and it can be identically recast in the form
\begin{equation}
{\cal J}^{\rm grav}[\xi] = d\left(\xi^\alpha H_\alpha + 
\Xi_\alpha{}^\beta
H^\alpha{}_\beta\right) + \xi^\alpha {\cal E}_\alpha + 
\Xi_\alpha{}^\beta
\,{\cal C}^\alpha{}_\beta,\label{JYd}
\end{equation}
where now $\Xi_\alpha{}^\beta =\xi\rfloor\Gamma_\alpha{}^\beta + 
\Theta_\alpha{}^\beta$.

\subsection{Matter current}

In an analogous way, if we consider the case when $V^{\rm tot} = L$,
we can derive the current for  matter. Usually, matter is described by
Lorentz-covariant (scalar or spinor) fields that are coupled to 
gravity in a general coordinate- and local Lorentz-covariant
manner (in accordance with the covariance and equivalence 
principles). Taking this into account, we conclude that the matter 
Lagrangian is constructed only from the covariant objects, i.e., $L = 
L(\Psi,D\Psi,T,R)$. Thus, the matter Lagrangian is invariant under 
the local Lorentz group, and the matter current can be written as in 
\cite{invar}:
\begin{equation}\label{Jmat1}
{\cal J}^{\rm mat}[\xi] := \xi\rfloor L - {\cal 
L}_\xi\vartheta^\alpha
\wedge {\frac {\partial L}{\partial T^\alpha}} - {\cal 
L}_\xi\Gamma_\alpha
{}^\beta\wedge {\frac {\partial L}{\partial R_\alpha{}^\beta}} -
{\cal L}_\xi\Psi^A\wedge {\frac {\partial L}{\partial D\Psi^A}}.
\end{equation}
This current satisfies
\begin{equation}
d{\cal J}^{\rm mat}[\xi]= {\cal 
L}_\xi\vartheta^\alpha\wedge\Sigma_\alpha
+ {\cal L}_\xi\Gamma_\alpha{}^\beta\wedge \tau^\alpha{}_\beta +
{\cal L}_\xi\Psi^A\wedge \frac{\delta L}{\delta\Psi^A}.\label{dJm}
\end{equation}
Furthermore, one can verify that
\begin{equation}\label{Jmat}
{\cal J}^{\rm mat}[\xi] = -\,d\left(\xi^\alpha{\frac {\partial L}
{\partial T^\alpha}} + \Xi_\alpha{}^\beta {\frac {\partial L}
{\partial R_\alpha{}^\beta}}+\xi\rfloor\Psi^A\wedge\frac{\partial
L}{\partial\Psi^A}\right) + \xi^\alpha\Sigma_\alpha
+ \Xi_\alpha{}^\beta\tau^\alpha{}_\beta 
+\xi\rfloor\Psi^A\wedge\frac{\delta
L}{\delta\Psi^A}.
\end{equation}

\subsection{Total current}

Let us now consider the total Lagrangian $V^{\rm tot} = V + L$
of the interacting gravitational and matter fields. The total current
is then the sum of the gravitational and the matter currents, given
respectively by eqs. (\ref{JYdef}) and (\ref{Jmat1}):
\begin{eqnarray}
{\cal J}[\xi]&=&{\cal J}^{\rm grav}[\xi] + {\cal J}^{\rm 
mat}[\xi]\nonumber\\
&=& \xi\rfloor (V + L) - \Theta^{\alpha\beta}v_{\alpha\beta} - 
{\cal L}_\xi\Psi^A\wedge {\frac {\partial L}{\partial D\Psi^A}} 
\nonumber\\ \label{Jd1}
&& + {\cal L}_\xi\vartheta^\alpha\wedge\left(H_\alpha - {\frac 
{\partial L} {\partial T^\alpha}}\right) + {\cal 
L}_\xi\Gamma_\alpha{}^\beta\wedge\left(H^\alpha{}_\beta 
- {\frac {\partial L}{\partial R_\alpha{}^\beta}}\right).
\end{eqnarray}
The total current, see (\ref{dJ}), satisfies
\begin{equation}
d{\cal J}[\xi]= {\cal L}_\xi\vartheta^{\alpha}\wedge ({\cal 
E}_\alpha + \Sigma_\alpha) + {\cal L}_\xi\Gamma_\alpha{}^{\beta}
\wedge({\cal C}^\alpha{}_\beta + \tau^\alpha{}_\beta) + 
{\cal L}_\xi\Psi^A\wedge{\frac{\delta L} {\delta\Psi^A}}.\label{dcurr}
\end{equation}
Accordingly, when the gravitational and matter fields satisfy 
the field equations (\ref{cFa})-(\ref{cFAA}), we obtain a 
conserved current: $d{\cal J}[\xi]= 0$. In addition, from 
(\ref{JYd}) and (\ref{Jmat}), or directly from (\ref{JdH}), we find
\begin{eqnarray}
{\cal J}[\xi] &=& d\left[\xi^\alpha\left(H_\alpha
- {\frac {\partial L}{\partial T^\alpha}}\right) + 
\Xi_\alpha{}^\beta
\left(H^\alpha{}_\beta - {\frac {\partial L}{\partial R_\alpha
{}^\beta}}\right)+\xi\rfloor\Psi^A\wedge\frac{\partial
L}{\partial\Psi^A}\right]\nonumber\\
&& +\,\xi^\alpha({\cal E}_\alpha + \Sigma_\alpha) + 
\Xi_\alpha{}^\beta
({\cal C}^\alpha{}_\beta + \tau^\alpha{}_\beta).\label{Jd}
\end{eqnarray}
Thus, the conserved total current is expressed in terms of the
superpotential 2-form ``on-shell". As a result, the conserved charge is
then computed as an integral over the spatial boundary $\partial S$:
\begin{equation}
{\cal Q}[\xi] := \int\limits_S {\cal J}[\xi] = 
\int\limits_{\partial S}\left[\xi^\alpha\left(H_\alpha - {\frac 
{\partial L}{\partial T^\alpha}}\right) + \Xi_\alpha{}^\beta\left(
H^\alpha{}_\beta - {\frac {\partial L}{\partial R_\alpha{}^\beta}}
\right)+\xi\rfloor\Psi^A\wedge\frac{\partial L}{\partial\Psi^A}\right].
\label{calq}
\end{equation}

This result generalizes the construction of the conserved currents
and charges, developed in \cite{invar}, to the case of the 
quasi-invariant models. Although the final formulas above and in 
\cite{invar} appear to be identical, there is an essential difference. 
For the theories with invariant Lagrangians, the gravitational field 
momenta are defined by $H_\alpha = - \partial V/\partial T^\alpha$ 
and $H^\alpha{}_\beta = - \partial V/\partial R_\alpha{}^\beta$. 
They are, by construction, {\it covariant} under local Lorentz 
transformations. This fact then guarantees that the conserved current 
and charge are true scalars under both diffeomorphisms and the local 
Lorentz group. In contrast, for the quasi-invariant models under 
consideration, the gravitational field momenta (\ref{HH}) are no 
longer covariant under local Lorentz transformations, not even for 
the choice of the Lie derivatives in the sense of Yano. As a result, 
both the conserved current (\ref{Jd1}) and the corresponding charge 
(\ref{calq}) are invariant under diffeomorphisms, but neither of them 
is a scalar under the local Lorentz group. This fact obviously 
represents a problem for the physical interpretation of the resulting 
conserved quantities.

\section{Tetrad formulation of gravity theory}

In order to demonstrate how the general formalism works in
physically interesting situations, we will now analyze the
4-dimensional teleparallel gravity theory in the so-called pure 
tetrad formulation. This model is described by the Lagrangian 
\cite{conserved}:
\begin{equation}
\tilde{V}(\vartheta,d\vartheta) = 
-\,\frac{1}{2\kappa}\,F^\alpha\wedge{}^\star
\left({}^{(1)}F_{\alpha} - 2\,{}^{(2)}\!F_{\alpha} 
-\frac{1}{2}\,{}^{(3)}\!F_{\alpha}\right). \label{V1}
\end{equation}
Here $\kappa=8\pi G/c^3$, and ${}^\star$ denotes the Hodge dual 
defined by the Minkowski metric $g_{\alpha\beta} = o_{\alpha\beta} 
:={\rm diag}(+1,-1,-1,-1)$. The latter is used also to raise and lower 
the Greek (local frame) indices. The 2-form $F^\alpha := d\vartheta^\alpha$ 
is the translational gauge field strength (which in geometric terms is equal 
to the anholonomity object of the tetrad). The irreducible decomposition of 
the field strength reads (see \cite{Gronwald97,HSh79,Itin,telemag} for details)
\begin{eqnarray}\label{Fi1}
{}^{(1)}F^{\alpha}&:=&F^{\alpha}-{}^{(2)}F^{\alpha}-{}^{(3)}F^{\alpha},\\
{}^{(2)}F^{\alpha}&:=&\frac{1}{3}\,\vartheta^\alpha\wedge\left(e_\beta\rfloor
F^\beta\right),\label{Fi2}\\
{}^{(3)}F^{\alpha}&:=&\frac{1}{3}\,e^\alpha\rfloor\left(\vartheta^\beta\wedge
F_\beta\right).\label{Fi3}
\end{eqnarray}
The variation of the total Lagrangian $V^{\rm tot}=\tilde{V} + L$ with
respect to the tetrad yields the gravitational field equations 
\begin{equation}
d\tilde{H}_\alpha - \tilde{E}_\alpha = \Sigma_\alpha. \label{fe1}
\end{equation}
Here, in accordance with the general Lagrange-Noether scheme (see
Sec.~\ref{LN}, and Refs. \cite{conserved,Gronwald97,HMMN95}), we have
\begin{eqnarray}
\tilde{H}_{\alpha} = -\,{\frac {\partial \tilde{V}} {\partial 
F^\alpha}} &=& {\frac{1}{2\kappa}}\tilde{\Gamma}^{\beta\gamma}
\wedge\eta_{\alpha\beta\gamma},\label{defH} \\
\tilde{E}_\alpha = {\frac {\partial \tilde{V}} {\partial 
\vartheta^\alpha}}
&=& e_\alpha\rfloor \tilde{V} + (e_\alpha\rfloor F^\beta)\wedge
\tilde{H}_\beta. \label{defE}
\end{eqnarray}
The teleparallel model (\ref{V1}) belongs to the class of 
quasi-invariant theories. In fact, one can verify that under a change of the 
coframe $\vartheta'^\alpha= \Lambda^\alpha_{\ \beta}(x)\,
\vartheta^{\beta}$, the Lagrangian changes by a total derivative:
\begin{equation}
\tilde{V}(\vartheta') = \tilde{V}(\vartheta) - {\frac 1 {2\kappa}}
\,d\left[(\Lambda^{-1})^\beta{}_\gamma 
d\Lambda^\gamma{}_\alpha\wedge
\eta^\alpha{}_\beta\right]. \label{Vprime}
\end{equation}
Hence, for this model we explicitly find $v^\alpha{}_\beta = - 
{\frac 1{2\kappa}}\,d\eta^\alpha{}_\beta$. The field equations 
(\ref{fe1}) are, however, Lorentz-covariant. As is well known, 
they coincide with the usual Einstein equations of general relativity 
theory. For this reason, the model (\ref{V1}) is usually called the  
teleparallel equivalent of general relativity.

In \cite{conserved}, we studied the transformation laws of the 
main objects in the tetrad formulation of gravity. In addition to 
(\ref{Vprime}), one can verify that
\begin{eqnarray}
\tilde{E}'_\alpha(\vartheta') &=&  (\Lambda^{-1})^\beta{}_\alpha
\tilde{E}_\beta(\vartheta) + d(\Lambda^{-1})^\beta{}_\alpha\wedge
\tilde{H}_\beta  \label{transE}\nonumber\\
&&\qquad -\,{\frac 1 {2\kappa}}\,d\left[(\Lambda^{-1})^\beta{}_\alpha
(\Lambda^{-1})^\nu{}_\gamma d\Lambda^\gamma{}_\mu\wedge\eta_\beta
{}^\mu{}_\nu\right], \label{Eprime}\\
\tilde{H}'_\alpha(\vartheta') &=&  (\Lambda^{-1})^\beta{}_\alpha
\tilde{H}_\beta(\vartheta) - {\frac 1 {2\kappa}}
(\Lambda^{-1})^\beta{}_\alpha (\Lambda^{-1})^\nu{}_\gamma
d\Lambda^\gamma{}_\mu\wedge\eta_\beta{}^\mu{}_\nu\,. \label{Hprime}
\end{eqnarray}

\subsection{Conserved charge in the tetrad gravity}

Using the general results of the previous section, we can write down
the conserved charge in the tetrad gravity theory as
\begin{equation}\label{Qtet1}
\tilde{\cal Q}[\xi,\vartheta] = \int\limits_{\partial S} \xi^\alpha
\tilde{H}_\alpha=\frac{1}{2\kappa}\int\limits_{\partial S}
\xi^\alpha\,\tilde{\Gamma}^{\beta\gamma}
\wedge\eta_{\alpha\beta\gamma} .
\end{equation}
This quantity substantially depends on the choice of the tetrad 
field configuration $\vartheta$. More exactly, under a local Lorentz
transformation, from (\ref{Hprime}) we find
\begin{equation}\label{Qtet2}
\widetilde{\cal Q}'[\xi,\vartheta'] = \widetilde{\cal
Q}[\xi,\vartheta] - \frac{1}{2\kappa}\int\limits_{\partial
S}\xi^\alpha (\Lambda^{-1})^\nu{}_\gamma
d\Lambda^\gamma{}_\mu\wedge\eta_\alpha{}^\mu{}_\nu.
\end{equation}
Note that the vector field $\xi$ does not depend on the choice 
of the frame, but its components $\xi^\alpha=\xi\rfloor\vartheta^\alpha$ 
transform as a vector.

\subsection{Charges for the Kerr solution}\label{examples}

Let us find the conserved charges for a particular tetrad 
configuration that describes an asymptotically flat axisymmetric 
rotating vacuum solution: the Kerr solution. We choose the Boyer-Lindquist
local coordinate system $(t,r,\theta,\varphi)$, 
and write the coframe as
\begin{eqnarray}
\vartheta^{\hat 0} &=& \sqrt{\frac{\Delta}{\Sigma}}\left[ 
c\,dt-a\sin^2\theta
\,d\varphi\right],\label{KNcof0} \\
\vartheta^{\hat 1} &=& \sqrt{\frac{\Sigma}{\Delta}}\, 
dr,\label{KNcof1} \\
\vartheta^{\hat 2} &=& \sqrt{\Sigma}\, d\theta,\label{KNcof2} \\
\vartheta^{\hat 3} &=& \frac{\sin\theta}{\sqrt{\Sigma}}\left[ 
-ac\,dt +(r^2+a^2)\,d\varphi\right], \label{KNcof3}
\end{eqnarray}
where the functions and constants are defined by
\begin{equation}
\Delta := r^2 + a^2 - 2mr,\quad
\Sigma := r^2 + a^2\cos^2\theta,\quad
m := \frac{GM}{c^2}.\label{DSm}
\end{equation}

Direct computation of the charge (\ref{Qtet1}) for the tetrad
(\ref{KNcof0})-(\ref{KNcof3}) yields a divergent result. 
However, we can choose another tetrad with the help of a suitable 
local Lorentz transformation of the original coframe.

Let us consider the Lorentz transformation described by the 
matrix $\Lambda =
\Lambda_1\Lambda_2\Lambda_3$, where
\begin{eqnarray}
\Lambda_1 &=& \left(\begin{array}{cccc} 1& 0 & 0 & 0\\ 0 & 
\cos\varphi & -\sin
\varphi & 0\\ 0 & \sin\varphi& \cos\varphi& 0\\ 0 & 0& 0& 
1\end{array}\right),
\label{L1}\\
\Lambda_2 &=& \left(\begin{array}{cccc} 1& 0& 0& 0\\ 0& 
(r/\sqrt{\Sigma})\sin
\theta& \sqrt{\Delta_0/\Sigma}\cos\theta & 0\\ 0& 0& 0& 1\\ 0& 
\sqrt{
\Delta_0/\Sigma}\cos\theta & -(r/\sqrt{\Sigma})\sin\theta &
0\end{array}\right),\label{L2}\\
\Lambda_3 &=& \left(\begin{array}{cccc} \sqrt{\Delta_0/\Sigma}& 
0& 0&
(a/\sqrt{\Sigma})\sin\theta\\ 0& 1& 0& 0\\ 0& 0& 1& 0\\ 
(a/\sqrt{\Sigma})
\sin\theta & 0& 0& 
\sqrt{\Delta_0/\Sigma}\end{array}\right).\label{L3}
\end{eqnarray}
Here $\Delta_0 = r^2 + a^2$. This Lorentz matrix defines a flat 
Lorentz connection $\overline{\Gamma}_\mu{}^\nu = (\Lambda^{-1})^\nu
{}_\gamma d\Lambda^\gamma{}_\mu$. Explicitly, we find:
\begin{eqnarray}
\overline{{\Gamma}}{}^{\hat{0}\hat{1}}&=& -\frac {ar \sin^2\theta}
{r^2+a^2\cos^2\theta}\,d\varphi,\qquad 
\overline{{\Gamma}}{}^{\hat{0}
\hat{2}}=-{\frac {a\,\sin \theta\cos\theta\sqrt {{r}^2+a^2}}
{r^2+a^2 \cos^2\theta}}\,d\varphi ,\label{G1}\\
\overline{{\Gamma}}{}^{\hat{0}\hat{3}}&=&-{\frac {ar\sin \theta}
{ \left( {r}^2+a^2 \cos^2\theta\right)\sqrt{{r}^2+a^2}}}\,dr+\frac{
a\cos\theta 
\sqrt{r^2+a^2}}{r^2+a^2\cos^2\theta}\,d\theta,\label{G2}\\
\overline{{\Gamma}}{}^{\hat{1}\hat{2}}&=&-\frac{a^2\sin\theta\cos\theta}
{(r^2+a^2 \cos^2\theta) \sqrt {{r}^2+a^2}}\,dr -\frac{r\sqrt{r^2+a^2}}
{r^2+a^2\cos^2\theta}\,d\theta ,\label{G3} \\ \label{G4}
\overline{{\Gamma}}{}^{\hat{1}\hat{3}}&=&-\frac{r\sin\theta\sqrt{r^2+a^2}}
{r^2+a^2\cos^2\theta}\,d\varphi,\qquad\overline{{\Gamma}}^{\hat{2}\hat{3}}
= -\frac{\left(r^2+a^2\right)\cos\theta}{r^2+a^2\cos^2\theta}\,d\varphi.
\end{eqnarray}

One can verify that the flat connection obtained in this way 
coincides with the Riemannian connection of the original tetrad 
(\ref{KNcof0})-(\ref{KNcof3}) in the limit of zero $m$, namely 
$\overline{\Gamma}_\mu{}^\nu = \tilde{\Gamma}_\mu{}^\nu\vline_{m=0}$. 
Using this observation, we can straightforwardly recast the charge 
(\ref{Qtet2}) computed for the transformed tetrad into the equivalent form
\begin{equation}\label{Qtet3}
\widetilde{\cal Q}'[\xi,\vartheta'] = \int\limits_{\partial S}
{\frac 1 {2\kappa}}\xi^\alpha\left(\tilde{\Gamma}^{\mu\nu} -
\overline{\Gamma}^{\mu\nu}\right)\wedge\eta_{\alpha\mu\nu}.
\end{equation}
This construction actually appears now to be ``invariant" under 
both diffeomorphism and local Lorentz transformations, as
$\left(\tilde{\Gamma}^{\mu\nu} -
\overline{\Gamma}^{\mu\nu}\right)$ is a covariant quantity. 
However, the integral 
obtained is only a result of the choice of a specific frame. Another 
frame will produce different values for the conserved quantities, in 
general, since it will correspond to different $\overline{\Gamma}$.

The direct evaluation of this integral, for vector fields $\xi$ with
constant components $\xi^i$ in the coordinate system used in
(\ref{KNcof0})-(\ref{KNcof3}), yields finite total conserved charges:
\begin{equation}\label{Qtet4}
\widetilde{\cal Q}'[\xi,\vartheta'] = \xi^0\,Mc^2 - 
\xi^3\,{\frac 23}Mca\,.
\end{equation}

\subsection{On the choice of the frame}

As we see, for the tetrad $\vartheta'^\alpha =
\Lambda^\alpha{}_\beta\vartheta^\beta$ determined by 
(\ref{L1})-(\ref{L3}), the conserved charge corresponding
to the diffeomorphism generated by the shift along the time 
coordinate has the usual value $\widetilde{\cal Q}'[\partial_t,
\vartheta'] = Mc^2$ of the total energy of the configuration. 
On the other hand, for the vector field along the azimuthal 
coordinate, we find $\widetilde{\cal Q}'[\partial_\varphi,
\vartheta'] = - {\frac 23}Mca$. This is proportional to the 
standard value of the total angular momentum of the Kerr solution, 
with the coefficient $2/3$. In a certain sense, the situation 
resembles the well-known outcome of the computation of the total 
mass and angular momentum by using the Komar formulas 
\cite{Komar1,Komar2} (although in that case the mass appears
with a ``wrong" coefficient $1/2$, and the angular momentum has 
the standard value $-Mca$). More important, however, is the fact 
that the total mass and angular momentum are both finite for this 
appropriate choice of the tetrad.

This is analogous to the well-known fact that the computation of 
the total energy and angular momentum in the framework of the 
energy-momentum pseudotensor technique necessarily requires a 
choice of a special class of spacetime coordinates. In the pure 
tetrad teleparallel approach, the results do not depend on the 
spacetime coordinates due to use of the explicitly diffeomorphism 
invariant formalism of exterior forms. However, the choice of 
the tetrad at spatial infinity becomes an essential aspect of
the computation of the conserved quantities.

This was manifested in the above calculation of the conserved 
charges for the asymptotically flat Kerr configuration. As we have 
demonstrated, in order to find physically acceptable finite values 
for the total energy and angular momentum, one needs to select a 
coframe in an appropriate way. The resulting coframe can be 
written as follows:
\begin{equation}
\vartheta'^\alpha = d\chi^\alpha
+ \left(\sqrt{\Delta} - \sqrt{\Delta_0}\right)\phi^\alpha.
\end{equation}
Here we denoted the quartet of functions
\begin{equation}
\chi^{\hat 0} = ct,\quad  \chi^{\hat 1} = \sqrt{\Delta_0}\sin\theta
\cos\varphi,\quad \chi^{\hat 2} = 
\sqrt{\Delta_0}\sin\theta\sin\varphi,
\quad \chi^{\hat 3} = r\cos\theta,
\end{equation}
and introduced the 1-form with the components
\begin{eqnarray}
\phi^{\hat 0} &=& {\frac {\sqrt{\Delta_0}}\Sigma}\,\zeta,\\
\phi^{\hat 1} &=& \sin\theta\left(-\,{\frac 
{r\cos\varphi}{\sqrt{\Delta
\Delta_0}}}\,dr - {\frac {a\sin\varphi}{\Sigma}}\,\zeta\right),\\
\phi^{\hat 2} &=& \sin\theta\left(-\,{\frac 
{r\sin\varphi}{\sqrt{\Delta
\Delta_0}}}\,dr + {\frac {a\cos\varphi}{\Sigma}}\,\zeta\right),\\
\phi^{\hat 3} &=& -\,{\frac {\cos\theta}{\sqrt{\Delta}}}\,dr,
\end{eqnarray}
where $\zeta = c\,dt - a\sin^2\theta\,d\varphi$. When $m=0$, 
this tetrad becomes holonomic, $\vartheta'^\alpha\vline_{m=0} = 
d\chi^\alpha$. In a sense, we may say that the transformed frame 
$\vartheta'$ describes the ``true'' gravitational field because, 
by ``switching off" the essential physical parameter $m$, the 
corresponding field strength $F'^\alpha = d\vartheta'^\alpha$
vanishes. In contrast, the 2-form $F^\alpha$ does not vanish 
with $m=0$ for the original coframe (\ref{KNcof0})-(\ref{KNcof3}), 
which manifests the mixed inertial-gravitational nature of this 
tetrad. Another appealing property of $\vartheta'$ is that it 
is asymptotically holonomic (at spatial infinity), and ---what 
is more important--- that its Riemannian connection
$\tilde{\Gamma}'(\vartheta')$ vanishes at spatial infinity. Exactly
this vanishing of the Riemannian connection makes the convergence 
of the integral of the conserved charge possible. This completely 
agrees with our recent observations for the conserved quantities 
discussed in \cite{conserved}.

Summarizing, the above analysis shows that among all possible 
tetrads that are formally allowed in teleparallel gravity, there 
exists a class of tetrads which are characterized, for asymptotically
flat gravitational field configurations, by the following properties: 
they (i)  have asymptotically 
vanishing Riemannian connection, (ii) the total conserved quantities 
(mass and the angular momentum) are finite for this class of frames. 
Notice that there remains a freedom to rotate these tetrads by means of 
a global Lorentz transformation. More generally, as the choice of the 
tetrad is relevant to the conserved quantities only at the boundary 
$\partial S$, it is even possible to choose different tetrads, related 
by a local Lorentz transformation $\Lambda^\alpha{}_\beta$ with 
$d\Lambda^\alpha{}_\beta\rightarrow 0$ on $\partial S$ such that the 
second term on the r.h.s. of (\ref{Qtet2}) is finite (or vanishes).

In the next section, we compare the results obtained within the 
purely tetrad (quasi-invariant) formulation with an alternative 
(explicitly invariant) approach to the teleparallel gravity based 
on the Poincar\'e gauge theory.

\section{Teleparallel model as a Poincar\'e gravity with constraints}

Teleparallel gravity can be naturally defined as a particular 
case of Poincar\'e gauge gravity \cite{telemag}. Its Lagrangian reads
\begin{equation} \label{V2}
V(\vartheta,d\vartheta,\Gamma,d\Gamma,\lambda) = -\,{\frac 1
{2\kappa}}T^{\alpha}\wedge{}^\star\left({}^{(1)}T_{\alpha}
- 2{}^{(2)}T_{\alpha} -{1\over 2}{}^{(3)}T_{\alpha}\right) 
- \lambda^\alpha{}_\beta\wedge R_\alpha{}^\beta.
\end{equation}
The last term imposes the vanishing curvature constraint, 
$R_\alpha{}^\beta
= d\Gamma_\alpha{}^\beta + 
\Gamma_\gamma{}^\beta\wedge\Gamma_\alpha{}^\gamma
= 0$, by means of the Lagrange multiplier 2-form 
$\lambda^\alpha{}_\beta$
(which is antisymmetric in its indices, $\lambda_{\alpha\beta} =
-\,\lambda_{\alpha\beta}$). An interesting feature of this model 
is that besides the diffeomorphism and local Lorentz symmetry, the 
action is also invariant under the transformation of the Lagrange 
multiplier
\begin{equation}
\lambda^\alpha{}_\beta\longrightarrow\lambda^\alpha{}_\beta + 
D\chi^\alpha{}_\beta,\label{lam}
\end{equation}
with an arbitrary 1-form $\chi^\alpha{}_\beta$. This is a direct 
consequence of the Bianchi identity $DR_\alpha{}^\beta \equiv 0$.

For the Lagrangian (\ref{V2}) we  find \cite{conserved,invar,PGrev}
\begin{equation}
H_\alpha = {\frac 1{2\kappa}}\,K^{\mu\nu}\wedge\eta_{\alpha
\mu\nu},\quad H^\alpha{}_\beta = \lambda^\alpha{}_\beta,\quad
E_\alpha = e_\alpha\rfloor V + (e_\alpha\rfloor T^\beta)\wedge 
H_\beta +e_\alpha\rfloor R_\gamma{}^\beta\wedge H^\gamma{}_\beta.
\end{equation}
Here, $K^{\mu\nu}$ is the contortion 1-form, given by the 
difference between the Riemannian (Christoffel) and dynamical 
Riemann-Cartan connections:
\begin{equation}
K_\alpha{}^\beta = \widetilde{\Gamma}_\alpha{}^\beta
- \Gamma_\alpha{}^\beta.\label{contor}
\end{equation}
The vacuum field equations derived from (\ref{V2}) by means of 
the variation w.r.t. $\vartheta^\alpha, \Gamma_\alpha{}^\beta$ 
and $\lambda^\alpha{}_\beta$ read, respectively:
\begin{eqnarray}
DH_\alpha - E_\alpha = {\frac 
1{2\kappa}}\,\widetilde{R}^{\mu\nu}\wedge
\eta_{\alpha\mu\nu} &=& 0,\label{1st}\\
DH_{\alpha\beta} + \vartheta_{[\alpha}\wedge H_{\beta]} \equiv
D\left(\lambda_{\alpha\beta} - {\frac 1{2\kappa}}\,\eta_{\alpha
\beta}\right) &=& 0,\label{2nd}\\
R_\alpha{}^\beta &=& 0.\label{R0}
\end{eqnarray}
The first equation (\ref{1st}) is the usual Einstein equation 
that determines the coframe (modulo local Lorentz transformations). 
{}From (\ref{2nd}) and (\ref{R0}) we find the Lagrange multiplier 
2-form
\begin{equation}
\lambda_{\alpha\beta} = {\frac 1{2\kappa}}\left(\eta_{\alpha\beta} +
D\chi_{\alpha\beta}\right).
\end{equation}
The 1-form $\chi_{\alpha\beta}$ is arbitrary, reflecting the 
gauge freedom (\ref{lam}). Finally, the constraint equation 
(\ref{R0}) determines the Riemann-Cartan connection as a flat 
connection.

The dynamics of this model is rather degenerate. Besides the 
freedom of the choice of the Lagrange multiplier, represented by 
(\ref{lam}), the equations for the coframe (tetrad) and for the 
connection, (\ref{1st}) and (\ref{R0}), are completely {\it uncoupled}. 
As a result, the flat Riemann-Cartan (i.e., Weitzenb\"ock) connection 
can be chosen in an arbitrary way, irrespectively of the value of 
the coframe. Mathematically this is manifested in the
possibility of performing {\it independent} local Lorentz 
transformations of the field variables:
\begin{equation}
\vartheta'^\alpha = \Lambda_I{}^\alpha_{\ 
\beta}\vartheta^\beta,\qquad
\Gamma_\alpha^{\prime\ \beta} = (\Lambda_{II}^{-1})^\mu_{\ 
\alpha}\Gamma_\mu{}^\nu\Lambda_{II}{}^\beta_{\ \nu} + 
\Lambda_{II}{}^\beta_{\ \gamma}
d(\Lambda_{II}^{-1})^\gamma_{\ \alpha},\label{addtrans}
\end{equation}
with two {\it different} Lorentz matrices $\Lambda_I{}^\alpha_{\ 
\beta}\neq \Lambda_{II}{}^\alpha_{\ \beta}$. As a result, we can 
``rotate" each of the two variables, either $\vartheta^\alpha$ 
or $\Gamma_\alpha{}^\beta$, while keeping another one fixed.

\subsection{Conserved invariant charge}

By using the above derivations, the invariant conserved charge 
for the teleparallel model is computed straightforwardly:
\begin{equation}
{\cal Q}[\xi] = {\frac 1{2\kappa}}\int\limits_S \left[\xi^\alpha 
K^{\mu\nu}\wedge\eta_{\alpha\mu\nu} + \Xi^{\alpha\beta}
\eta_{\alpha\beta} - ({\cal L}_\xi\Gamma^{\alpha\beta})
\wedge\chi_{\alpha\beta}\right].
\end{equation}
Here $\Xi_\alpha{}^\beta =\xi\rfloor\Gamma_\alpha{}^\beta + 
\Theta_\alpha{}^\beta$, see (\ref{Theta}). We have also used 
the identity (A13) of \cite{invar}. Recalling that the 1-form 
$\chi_{\alpha\beta}$ is completely undetermined by the field 
equations, we conclude that a unique conserved charge can only
be defined for {\it symmetric} field configurations that satisfy 
the generalized Killing equation ${\cal L}_\xi\Gamma^{\alpha\beta} =0$.

Using the identities (A18) of \cite{invar} it is direct to prove 
that, for solutions with vanishing torsion, the Killing equation 
${\cal L}_\xi\vartheta^\alpha=0$ implies also that ${\cal L}_\xi
\Gamma_\alpha{}^\beta=0$. Therefore, for the Kerr solution we have 
${\cal L}_\xi\Gamma_\alpha{}^\beta=0$ for $\xi=\xi^0\partial_t
+\xi^3\partial_\varphi$, with constant $\xi^0$ and $\xi^3$.
As a result, for such symmetric configurations, the undetermined 
piece of the Lagrange multiplier disappears from the integral, and 
the conserved charge reduces to
\begin{equation}
{\cal Q}[\xi] = {\frac 1{2\kappa}}\int\limits_{\partial S} \left[
\xi^\alpha\left(\tilde{\Gamma}^{\mu\nu} - \Gamma^{\mu\nu}\right)
\wedge\eta_{\alpha\mu\nu} + \Xi^{\alpha\beta}
\eta_{\alpha\beta}\right].\label{Qcon1}
\end{equation}
Here we used the definition of the contortion (\ref{contor}). It is
worthwhile to compare this formula with the conserved charge in the
pure tetrad formulation (\ref{Qtet3}). As we see, the first term
(\ref{Qcon1}) directly corresponds to (\ref{Qtet3}) with the 
background connection replaced by the dynamical one. There is, 
however, a second term that does not have counterparts in the 
pure tetrad framework.

\subsection{Charges for Killing vectors of the Kerr solution}

Let us now compare the computation of the conserved charges in 
the two frameworks (purely tetrad and Poincar\'e gravity with 
constraints) for a specific configuration. We consider again 
the Kerr solution. Since the conserved charges are well defined 
only for the generalized Killing vector fields, we now deal 
with these symmetric configurations.

One can directly check that the Kerr coframe (\ref{KNcof0})-(\ref{KNcof3})
satisfies the symmetry condition $\ell_\xi\vartheta^\alpha =0$ 
for the vector field $\xi = \xi^0\partial_t + \xi^3\partial_\varphi$, 
with constants $\xi^0$ and $\xi^3$. Now we have to choose the Weitzenb\"ock 
connection $\Gamma$. It is a flat Riemann-Cartan connection and hence it 
can always be constructed as $\Gamma_\mu{}^\nu = (\Lambda^{-1})^\nu
{}_\gamma d\Lambda^\gamma{}_\mu$ with some Lorentz matrix $\Lambda$. 
It is easy to check that the matrix $\Lambda = \Lambda_1\Lambda_2\Lambda_3$ 
with the factors defined by (\ref{L1})-(\ref{L3}) yields an appropriate
choice. The resulting components of the Weitzenb\"ock connection are
then given by the formulas (\ref{G1})-(\ref{G4}). Finally, we 
can substitute everything into the formula (\ref{Qcon1}) and the direct 
evaluation of the integrals over the spatial boundary gives the result:
\begin{equation}\label{Qcon2}
{\cal Q}[\xi] = \xi^0\,Mc^2 - \xi^3\,Mca\,.
\end{equation}
As we can see, the last term in (\ref{Qcon1}) turns out to be very important
since then the total energy and the total angular momentum have 
their standard values, ${\cal Q}[\partial_t] = Mc^2$ and ${\cal Q}
[\partial_\varphi] = - Mca$, improving the unusual coefficients in
(\ref{Qtet4}).

\subsection{An alternative Lagrangian for teleparallel gravity?}

The comparison of the purely tetrad formulation with the Poincar\'e
gauge theory with constraint reveals an interesting observation: 
both approaches are not ideal. Namely, in the purely tetrad 
formulation we work with the coframe $\vartheta^\alpha$ as the only 
dynamical variable. It is completely determined (up to local Lorentz 
rotations) by the field equations. However, the resulting conserved 
current and charge are not invariant under local Lorentz transformations. 
Moreover, in general the total conserved quantities are divergent. 
One should choose a tetrad field that satisfies certain conditions 
in order to obtain physically meaningful total charges. The situation 
in the Poincar\'e approach is in a certain sense complementary. Namely,
the conserved current and charge are explicitly invariant under both
diffeomorphisms and local Lorentz transformations. However, for their
computation one needs, besides the tetrad, to know the connection and 
the Lagrange multiplier. Both of these variables are not determined 
by the field equations in a unique local-Lorentz covariant way. 
The arbitrariness in the Lagrange multiplier can be avoided by imposing 
generalized symmetry conditions on the field configurations. But in the 
choice of the flat connection there still remains a freedom similar to 
the freedom of the choice of the coframe in the tetrad formulation.

It seems that one can avoid many (not all, though) difficulties
mentioned above by using a different dynamical scheme. Let us
outline it here briefly. We can consider the following
Lorentz-invariant Lagrangian:
\begin{equation} \label{V3}
V(\vartheta,d\vartheta,\Gamma,\lambda) = -\,{\frac 1 {2\kappa}}
T^{\alpha}\wedge{}^\star\left({}^{(1)}T_{\alpha}- 2{}^{(2)}T_{\alpha} 
-{1\over 2}{}^{(3)}T_{\alpha}\right) - \lambda^\alpha{}_\beta\wedge
\left(\Gamma_\alpha{}^\beta-\overline{\Gamma}_\alpha{}^\beta\right).
\end{equation}
Here $\lambda^\alpha{}_\beta$ is a Lagrange multiplier 3-form, 
which imposes the teleparallel constraint by making the connection to  
reduce to the flat ``background'' connection $\overline{\Gamma}$,
$R_\alpha{}^\beta(\overline{\Gamma})=0$.

The vacuum field equations derived from (\ref{V3}) by means of 
the variation w.r.t. $\vartheta^\alpha$, yields the Einstein 
equation (\ref{1st}), as before. Variations w.r.t. $\Gamma_\alpha{}^\beta$
and $\lambda^\alpha{}_\beta$ lead to
\begin{eqnarray}
\lambda_{\alpha\beta} + \frac{1}{2\kappa}\,D\eta_{\alpha\beta}
&=& 0,\label{2nda}\\
\Gamma_\alpha{}^\beta&=&\overline{\Gamma}_\alpha{}^\beta,\label{R0a}
\end{eqnarray}
respectively. The invariant conserved charge for the teleparallel model
(\ref{V3}) is given by: 
\begin{equation}
{\cal Q}[\xi,\vartheta,\overline{\Gamma}] = 
\frac{1}{2\kappa}\int\limits_S
\xi^\alpha\left(\Gamma^{\mu\nu}-\overline{\Gamma}^{\mu\nu}\right)
\wedge\eta_{\alpha\mu\nu},
\end{equation}
which is formally identical to (\ref{Qtet3}). Here ${\cal
Q}[\xi,\vartheta,\overline{\Gamma}]$ is invariant under local 
Lorentz transformations, but depends on the choice of the flat 
background connection $\overline{\Gamma}$. The choice of 
$\overline{\Gamma}$ is, however, equivalent to the choice of a 
preferred tetrad frame $\vartheta'$ in which $\overline{\Gamma}'=0$
and ${\cal Q}[\xi,\vartheta,\overline{\Gamma}]={\cal Q}[\xi,
\vartheta',0]$, which then reduces to (\ref{Qtet1}) for the frame 
$\vartheta'$, i.e., ${\cal Q}[\xi,\vartheta',0]=\tilde{\cal Q}[\xi,
\vartheta']$. As a consequence of these observations, for the frame 
$\vartheta$ defined by (\ref{KNcof0})-(\ref{KNcof3}) and the 
``background'' connection (\ref{G1})-(\ref{G4}), we obtain the 
finite values (\ref{Qtet4}).

\section{Discussion and conclusions}

Our results for the teleparallel gravity models can be 
interpreted as follows. Tetrad frames differing by a local Lorentz 
transformation are related to different Lorentz connections.
A connection includes both inertial and gravitational effects 
\cite{paris}. While the gravitational effects produced by compact 
sources vanish asymptotically, the inertial effects can grow up at 
large distances. This is the case, for example, of the inertial
effects in rotating frames. As a consequence, when calculated in 
a general tetrad frame, the noninvariant conserved quantities can 
diverge due to the inertial effects carried by that frame. In order
to get physically meaningful conserved charges, therefore, it is 
crucial to choose an appropriate (or preferred) frame, in which 
the inertial effects are absent, in the sense that the connection 
vanishes asymptotically. It should be noted that the coordinate 
counterpart of this property is well known. In fact, it has already 
been remarked by several authors that, in order to obtain a finite 
value for the energy of a gravitational system in the usual
pseudotensor approach, the integration must be carried out in an 
asymptotically Minkowskian coordinate system \cite{coordinate,vargas}. 
Similarly to the choice of the preferred frame, the choice of this 
coordinate system is crucial in the sense that it does not introduce 
spurious effects in the calculation of the energy.

The choice of an appropriate frame is ultimately equivalent to 
the choice of an appropriate connection. This is the idea behind 
the strategy of using a background connection. To understand this 
point, let us recall that the inertial effects are not covariant. 
This is quite clear in special relativity, where we know that the 
inertial effects are not present in the specific class of
inertial frames, but do appear in any other class of frames. Being
non-covariant, these effects turn out to be represented by a 
connection \cite{inertial}. Therefore, if $\Gamma_\alpha{}^\beta$ 
is the connection in a general (orthonormal) frame, before calculating 
the conserved charges, it is necessary to extract from it all inertial 
effects connected with the frame, which here are represented by the 
background connection $\overline{\Gamma}_\alpha{}^\beta$ \cite{maluf}. 
When we do that, the resulting conserved charge will represent purely 
gravitational effects, and can consequently be finite. A related open 
question is whether there exist a frame $\vartheta$ for which 
(\ref{Qtet1}) (or, equivalently, a connection $\overline{\Gamma}$ for 
which (\ref{Qtet3})) leads to the standard values ${\cal
Q}[\xi] =Mc^2$ and ${\cal Q}[\xi]=-Mca$ for the Kerr metric.

Summing up, we have developed a general formalism for  
constructing conserved currents and charges in the gravitational models 
with quasi-invariant Lagrangians. For such models, the conserved current
(\ref{Jd1}) and thus the charge (\ref{calq}) are not invariant under
local Lorentz transformations even when the field equations are 
covariant. As an application, we then analyzed an important case of the 
teleparallel gravity. Another interesting application seems to be 
the class of models in 3 (or 5 and higher odd dimensions) with topological 
terms included in the Lagrangian. The case of a 3-dimensional gravity 
theory of that type was earlier studied in \cite{Bak94}, where the 
corresponding conserved quantities were explicitly derived. The complete 
analysis of this class of models will be given elsewhere.

\begin{acknowledgments}
The authors would like to thank FAPESP (YNO and JGP), CNPq 
(GFR and JGP) and CAPES for financial support.
\end{acknowledgments}

\end{document}